\documentclass[11pt,a4paper]{article}
\usepackage[utf8]{inputenc}
\usepackage[english]{babel}
\usepackage{amsmath}
\usepackage{amsfonts}
\usepackage{amssymb}
\usepackage{graphicx}
\usepackage{placeins}
\usepackage{subcaption}
\usepackage[left=3cm,right=3cm,top=3cm,bottom=3cm]{geometry}
\usepackage{color,soul}
\usepackage{xcolor}
\usepackage{natbib}
\usepackage{mathtools}
\usepackage{setspace}
\usepackage{hyperref}
\usepackage{lineno}

\begin{document}

 
\author{T. Zerenner$^a$\footnote{t.zerenner@gmail.com}, F. Di Lauro$^{a,b}$, M. Dashti$^a$, L. Berthouze$^c$, I. Z. Kiss$^a$\footnote{i.z.kiss@sussex.ac.uk}}

\date{%
    $^a$\footnotesize \textit{Department of Mathematics, University of Sussex, Falmer, Brighton, BN1 9QH, UK\\%
    $^b$Big Data Institute, Nuffield Department of Medicine, University of
Oxford, Oxford, OX3 7FL, UK\\%
    $^c$Department of Informatics, University of Sussex, Falmer, Brighton, BN1 9QH, UK}\\[2ex]%
    \today
}
\title{Probabilistic predictions of SIS epidemics on networks based on population-level observations}

\maketitle

\begin{abstract}

    We predict the future course of ongoing susceptible-infected-susceptible (SIS) epidemics on regular, Erd\H{o}s-R\'{e}nyi and Barab\'asi-Albert networks. It is known that the contact network influences the spread of an epidemic within a population. Therefore, observations of an epidemic, in this case at the population-level, contain information about the underlying network. This information, in turn, is useful for predicting the future course of an ongoing epidemic. To exploit this in a prediction framework, the exact high-dimensional stochastic model of an SIS epidemic on a network is approximated by a lower-dimensional surrogate model. The surrogate model is based on a birth-and-death process; the effect of the underlying network is described by a parametric model for the birth rates. We demonstrate empirically that the surrogate model captures the intrinsic stochasticity of the epidemic once it reaches a point from which it will not die out. Bayesian parameter inference allows for uncertainty about the model parameters and the class of the underlying network to be incorporated directly into probabilistic predictions. An evaluation of a number of scenarios shows that in most cases the resulting prediction intervals adequately quantify the prediction uncertainty. 
    As long as the population-level data is available over a long-enough period, even if not sampled frequently, the model leads to excellent predictions where the underlying network is correctly identified and prediction uncertainty mainly reflects the intrinsic stochasticity of the spreading epidemic.
    For predictions inferred from shorter observational periods, uncertainty about parameters and network class dominate prediction uncertainty.  
    The proposed method relies on minimal data at population-level, which is always likely to be available. This, combined with its numerical efficiency, makes the proposed method attractive to be used either as a standalone inference and prediction scheme or in conjunction with other inference and/or predictive models.
   
\end{abstract}

    \section{Introduction}
        \label{sec:introduction}

Mathematical models of the dynamics of directly transmitted infectious diseases can provide predictions about the future course of an ongoing epidemic and hence aid in decision making and epidemic control \citep[e.g.][]{siettos2013mathematical}. Statistical time series methods are utilised predominantly for epidemic nowcasting, i.e., shortest-term predictions and current state assessment under not yet complete data \citep[e.g.][]{bastos2019modelling,mcgough2020nowcasting} and epidemic surveillance, i.e., early identification of emerging epidemics \citep{unkel2012statistical}. Mechanistic/state-space models, which are based on a mathematical description of the spreading process, allow one to make predictions about the future course of an epidemic \citep{shaman2012forecasting, tizzoni2012real, nsoesie2013forecasting} as well as simulating intervention strategies \citep{chao2011planning, di2021optimal, di2020impact, van2020covid}. Many such models rely on a compartmentalisation of a population of $N$ individuals according to the individual's disease status. In diseases for which there is no immunity upon recovery, each individual is either susceptible (S) or infected/infectious (I) at any given time.

One assumption common to many compartmental models is that of random mixing of individuals, or of/within subgroups of a population \citep[e.g.,][]{kermack1927contribution, jacquez1993stochastic}. While this assumption can be adequate in some instances \citep[e.g., within households;][]{goeyvaerts2018household}, it is known that populations do not mix at random in general. Rather individuals have a finite set of contacts to whom they can pass on an infection. It is well-established that the contact network of a population significantly impacts epidemic dynamics \citep[e.g.,][]{shirley2005impacts, yin2017impact}. The importance of contact structure for epidemic dynamics has led to a close interaction between network science and mathematical epidemiology \citep{keeling2005networks, danon2011networks, pastor2015epidemic, kiss2017mathematics} whereby the spread of an epidemic within a population is understood and modelled as a stochastic process on a network. In such a model, each individual in the population corresponds to a node in the network, and a contact that represents a potential route for disease transmission between two individuals is a link in the network.

Drawbacks of network epidemiological models typically include their high dimensionality and the inaccessibility of the exact contact network of a population. Consider a SIS-epidemic on an undirected, unweighted network with $N$ nodes. At any given time, each node is either susceptible or infected/infectious. If the exact contact network is static and known, a complete description of the SIS dynamics is given by a continuous-time Markov-chain of dimension $2^N$ \citep[one equation for each possible network state; e.g.,][]{simon2011exact}. Such a Markov-chain model is exact, but also high-dimensional even for modest values $N$. Hence, the numerical integration of the system of equations becomes unfeasible for most real-life networks. Consequently, analytical results based on the exact system are mostly out of reach, and existing results typically rely on mean-field approximations \cite[e.g.,][]{mata2013pair, cota2018robustness}. Further, the exact contact network of a population is rarely accessible, but usually needs to be approximated either from limited observations and/or based on theoretical network models \citep[e.g.,][]{della2020network, xue2020data}.

In this study, we explore the suitability of a computationally inexpensive model to describe the stochastic process of an SIS epidemic spreading on Regular (Reg), Erd\H{o}s–R\'{e}nyi (ER) and Bar\'abasi-Albert (BA) networks. The surrogate model utilised in this study was first introduced in~\cite{di2020network} and further expanded to include more network classes and consider the large $N$ limit in~\cite{di2020pde}. The core idea of the approach is a dimension reduction of the state space. In the surrogate model, the state of the epidemic at any given time is defined by the total number of infected nodes in the population. The effect of the contact structure on the spreading of the epidemic is accounted for by the model parameters. The continuous-time Markov-chain describing the SIS dynamics on the reduced state space takes the form of a Birth-and-Death (BD) process and is of dimension $N+1$; that is, it is linear in $N$ and thus feasible also for large $N$. The parameters of the BD model correspond to recovery and infections rates. The recovery rate is network-independent and here assumed to be known. The rate at which new infections occur depends on the network, but for particular network classes it can be well described by a three-parameter model, reducing the number of free parameters of the BD model from $2(N+1)$ to only three. \cite{di2020network} utilised the finding that different types of networks are associated with distinct regions in the space spanned by the three parameters to infer the type of network from population-level observations. To solve this inverse problem \cite{di2020network} set up a Bayesian inference procedure and built network class specific prior distributions which then allow to identify the most likely network class from the posterior.

Here, we utilise the BD model to forecast the evolution of an on-going epidemic. We address the following questions:

\begin{itemize}
    \item How well does the BD model capture the intrinsic stochasticity of an epidemic spreading on a network?
    \item How uncertain are model parameters when inferred from the kind of time-censored observations typically available in a realistic prediction scenario, and how does this uncertainty translate into prediction uncertainty?
    \item Can we use the BD model and Bayesian inference to provide epidemic forecasts with meaningful uncertainty information?
\end{itemize}

The manuscript is structured as follows: Section~\ref{sec:method} introduces the BD model and Bayesian inference. We then outline the generation and evaluation of predictions using the BD model. We consider nine different combinations of networks and epidemic parameters: a small, a medium and a large epidemic on a network from each of the three aforementioned network classes (Section~\ref{sec:data}). An empirical validation of the BD model based on these nine cases is provided in Section~\ref{sec:results1}. In Section~\ref{sec:results2}, we evaluate the predictions obtained with the BD model for all nine cases. In particular, we study the sensitivity of network class and parameter inference on the number and timing of observations in realistic prediction scenarios and how uncertainty about network class and model parameters translates into prediction uncertainty. We conclude with a discussion including limitations of this work and future directions.
    
    \section{Methods}
        \label{sec:method}


\subsection{SIS epidemics on networks}
\label{sec:SIS}

We consider the standard SIS epidemic on a population of $N$ individuals whose contact structure is described by an undirected and unweighted network defined by its adjacency matrix $G = (g_{ij})$ with $i,j=1,2, \dots N$ and $g_{ij}=1$ if individuals (nodes) $i$ and $j$ are connected and $g_{ij}=0$ otherwise. If two nodes $i$ and $j$ are connected, the disease can be transmitted from one to the other. In an SIS epidemic, each node is, at any given time, either susceptible (S) or infected/infectious (I). Thus, the epidemic state of the network at time $t$ is described by a Boolean vector $X(t) = (x_{i}(t))$ with $i=1,2, \dots N$ where $x_{i}(t)=0$ if node $i$ is susceptible and $x_i(t)=1$ if node $i$ is infected at time $t$. Hence, there exists a total of $2^N$ distinct network states. The state of the network changes through two types of events: the recovery or the infection of a node. Infection and recovery are Markovian and act as homogeneous Poisson point processes with constant per-link infection rate $\tau$ and constant recovery rate $\gamma$. An infectious node can spread the infection only to neighbouring susceptible nodes. Infection dynamics thus depends on the network structure, while recovery is network-independent. A complete description of the SIS-dynamics on a given network corresponds to a Markov-chain over a state-space of dimension $2^N$ for which numerical integration becomes intractable even for modest values of $N$. However, given network adjacency $G$, epidemic parameters $\tau$ and $\gamma$, and initial conditions $X_0=X(t_0)$, realisations of the stochastic process can be readily obtained using the Gillespie algorithm \citep[e.g.,][]{gillespie1977exact, kiss2017mathematics}. This is computationally comparatively inexpensive and provides us with i.i.d. samples of the true stochastic process. Such samples serve as a reference for the validation of the BD model and for the evaluation of predictions. More precisely, we make use of the aggregated number of infected nodes in the network, i.e., $I(t)=\sum_{i=1}^{N} x_i(t)$, which describes the epidemic at population level.


\subsection{BD model}
\label{sec:surrogatemodel}

Birth-and-death processes are intuitively linked to the population-level dynamics of SIS epidemics \cite[e.g.,][]{ganesh2005effect, nagy2014approximate, devriendt2017unified}. In this view, an increase in the number of infected individuals ($I\rightarrow I+1$) corresponds to the 'birth of an infection'; a decrease ($I\rightarrow I-1$) to the 'death of an infection'. Accordingly, the epidemic state is defined by the number of infected nodes $I \in \{0, \dots, N\}$ in the network. The resulting model is, like the exact formulation (Sec.~\ref{sec:SIS}), a continuous-time Markov chain, but on a state space of dimension $N+1$ only. 

\begin{figure}[t]
        \centering
        \includegraphics[width=\linewidth]{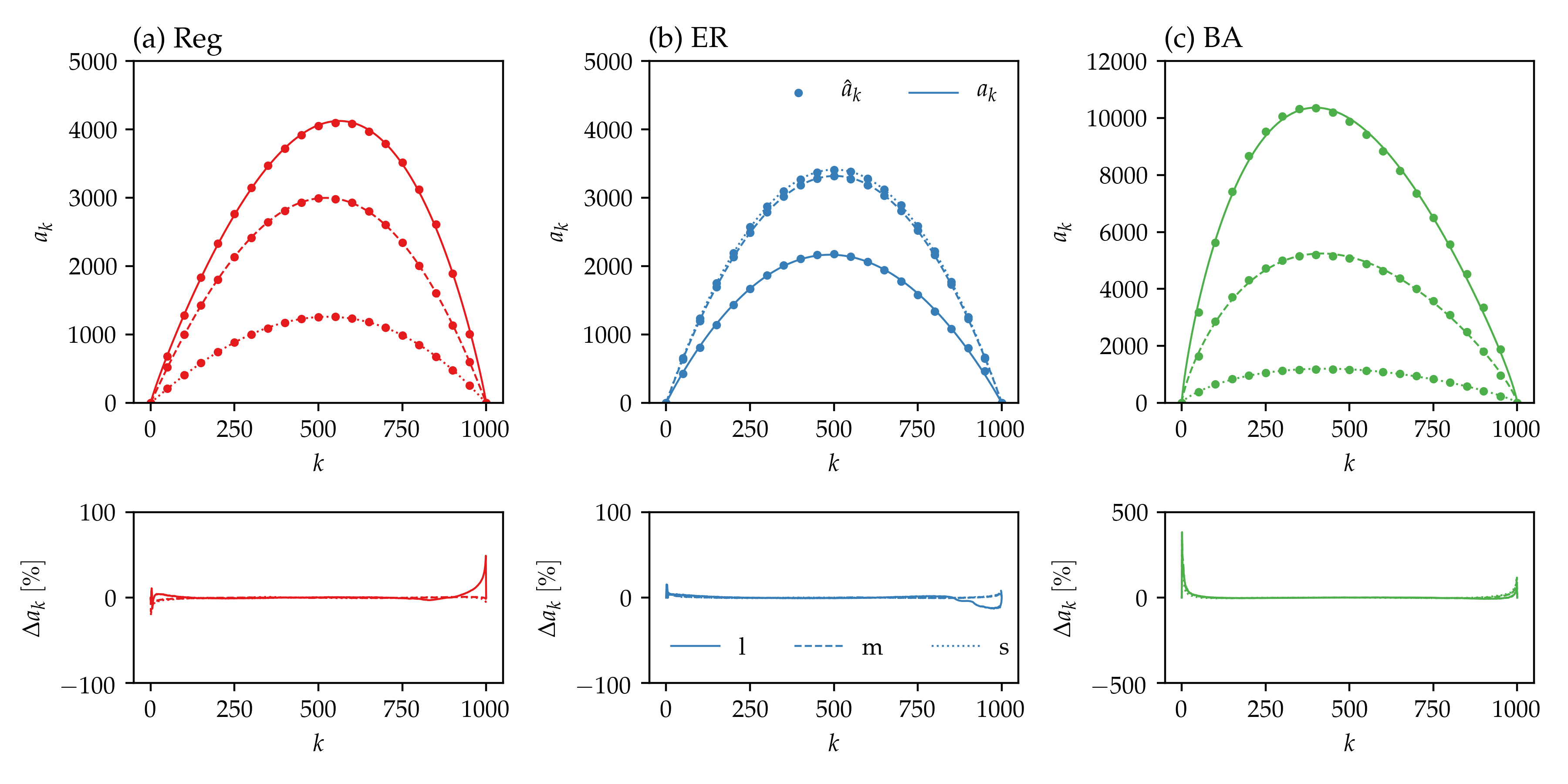}
        \caption{Parametric $a_k(C,\alpha,p)$ model fitted to $\hat{a}_k$ from Gillespie simulations of large (l), medium (m) and small (s) epidemics on the different networks. In the top panels, the dots indicate $\hat{a}_k$ and the lines correspond to $a_k(C,\alpha,p)$ with parameters $(C,\alpha,p)$ obtained from a least-squares fit of Eq.~\ref{eq:cap} to $\hat{a}_k$. The bottom panels show the relative error $\Delta a_k = (a_k(C,\alpha,p)-\hat{a}_k)/(\hat{a}_k+1)\times 100$. (Adding one in the denominator allows to also include the values at and in the vicinity of $k=0$ and $k=1000$.)}
        \label{fig:ak_curves}
    \end{figure}

The Kolmogorov (or Master) equation of a standard BD process is given by

\begin{equation}
    \forall k \in \{0,\dots,N\},~\dot{p}_{k}(t) = a_{k-1}~p_{k-1}(t) - (a_k + c_k)~p_{k}(t) +  c_{k+1}~p_{k+1}(t),
    \label{eq:BD}
\end{equation}

where $p_{k}(t)$ denotes the probability of observing $k$ infected nodes at time $t$, and $a_k$ and $c_k$ denote population-level infection and recovery rate, respectively. We note that $a_{-1} = c_{N+1} = 0$. The population-level recovery rate $c_k$ can be directly obtained from the node recovery rate $\gamma$ as $c_k = \gamma k$. The population-level infection rate $a_k$ however depends on the number of links between susceptible and infected nodes (S-I links) present in the network in its current state and is thus a random variable depending on the precise network. Following \cite{di2020network}, $a_k$ is represented in the BD model by its expectation $\hat{a}_k = \tau \times$\textit{the time-averaged number of S-I links over the network states with $k$ infected nodes}. \cite{di2020network} further demonstrated that for Regular, Erd\H os–R\'{e}nyi and Barab\'asi-Albert networks $\hat{a}_k$ can be well represented by a three-parameter model of the form

    \begin{equation}
        \forall k \in \{0,\dots,N\},~a_k(C,\alpha,p) = C~k^p~(N-k)^p \left( \alpha \left( k-\frac{N}{2}\right)+N \right),
        \label{eq:cap}
    \end{equation}
    
with $C$ serving as a general scaling parameter, $\alpha$ allowing to shift the peak of the curve with respect to $k=N/2$ and $p$ adjusting the flatness of the curve. The shape of the $a_k$ curves is network class-specific (Fig.~\ref{fig:ak_curves}). Whilst the peak is located near the centre ($k=N/2$) for Erd\H{o}s–R\'{e}nyi networks, it is shifted to the right for Regular networks, and to the left for Barab\'asi-Albert networks. Accordingly, the  $(C,\alpha,p)$-triplets for the different network types cluster in different regions in the three-dimensional parameter space. This observation is central to the network class inference of \cite{di2020network}.

For a given $\gamma$, a given $(C,\alpha,p)$-triplet, and initial conditions $p_k(t_0)=1$ if $k=I(t_0)$ and $p_k(t_0)=0$ otherwise, where $I(t_0) \in \{0,\dots,N\}$ denotes the number of infected nodes at $t_0$, we can numerically integrate Eq.~\ref{eq:BD} to obtain predictions $p_k(t)$. In our experiments, we assume that the recovery rate $\gamma$ is known. Initial conditions are provided by the last observation available. Thus, the remaining task is the inference of $(C,\alpha,p)$ from the available observational data, here, the number of infected nodes at a set of discrete times. We use Bayesian inference to estimate the posterior distribution over the parameters $(C,\alpha,p)$ given observations. The required priors were derived from extensive Gillespie simulations on different networks and with different epidemic parameters. The particular challenge for making predictions lies in the limited observational period available for inference in a realistic prediction scenario, in which observations exist only up to the current state of the epidemic.


\subsubsection{Bayesian inference}

The detail of the inference framework can be found in \cite{di2020network}. Here, we only recall the main ideas and steps of the inference procedure. We denote the population level observations by $(y,s)$ where $y=(k_1,....,k_n)$ with $k_j \in \{0,...,N\}$ denotes the number of infected individuals at times $s=(t_1,....,t_n)$. For brevity, we use $u$ to denote $(C,\alpha,p)$. We further denote the set of candidate network classes as $\Theta = \{\mathrm{Reg,~ER,~BA}\}$.

In order to make predictions, we require the posterior over $u$ given the observations, i.e., $\pi(u|y,s)$, which we can write as

\begin{equation*}
\pi(u|y,s) = \sum_{\Theta} \pi_{\theta}(u|y,s) ~ \pi(\theta|y,s).
\end{equation*}

In \cite{di2020network}, the goal was network class inference, i.e., obtaining the posterior over $\Theta$ given observations $(y,s)$, $\pi(\theta|y,s)$. To this end, \cite{di2020network} generated network class specific priors $\pi_{0,\theta}(u)$. Precisely, they carried out a large number of Gillespie simulations during which they kept track of the number of infected nodes $k$, the number of S-I links in the respective network states as well as the time spent in the various states. The parametric $a_k(C,\alpha,p)$ model from Eq.~\ref{eq:cap} was then fitted to the $(k,\hat{a}_k)$ curves from the Gillespie simulations by a least-squares fit using a particle swarm algorithm \citep{kennedy1995particle}. The resulting $(C,\alpha,p)$ triplets were used to infer Gaussian kernel density estimators \citep{pedregosa2011scikit} for the priors $\pi_{0,\theta}(u)$. 
Assuming a non informative, uniform prior for network class $\theta$, the prior distribution over $\theta$ and $u$ is given by

\begin{equation*}
    \pi_0(u,\theta) = \frac{1}{3} \pi_{0,\theta}(u).
\end{equation*}

Employing Bayes' rule we obtain the network class specific posterior(s) over the parameter space as

\begin{equation} \label{eq:posterior0}
\begin{split}
\pi_{\theta}(u|y,s) & \propto \mathcal{L}^u(y,s)~\pi_{0,\theta}(u).
\end{split}
\end{equation}

and the posterior over the network classes as 

\begin{equation*} \label{eq1}
\begin{split}
\pi(\theta|y,s) & =  \int \pi(u,\theta|y,s) du \\
                & \propto \int \mathcal{L}^u(y,s) \pi_{0,\theta}(u) du,
\end{split}
\end{equation*}

where $\mathcal{L}^u(y,s)$ denotes the likelihood of the observations under the forward model from Eq.~\ref{eq:BD} which is given by

\begin{equation}
   \mathcal{L}^u(y,s) = \prod_{i=1}^{n-1} p^u_{k_i,k_{i+1}}(t_{i+1} - t_i).
   \label{eq:likelihood}
\end{equation}

Following \cite{di2020network}, the terms $p^u_{k_i,k_{i+1}}$ are computed using the algorithm from \cite{crawford2014estimation}. The Python implementation routine estimating $\pi(\theta|y,s)$ is available at \url{https://github.com/BayIAnet/NetworkInferenceFromPopulationLevelData}. To estimate $\pi_{\theta}(u|y,s)$, we draw samples from $\pi_{\theta}(u|y,s)$ using the Metropolis–Hastings algorithm, making use of Eqs.~\ref{eq:posterior0} and~\ref{eq:likelihood}.
        \subsubsection{Prediction and uncertainty}

\begin{figure}[t]
     \centering
     \includegraphics[width=\linewidth]{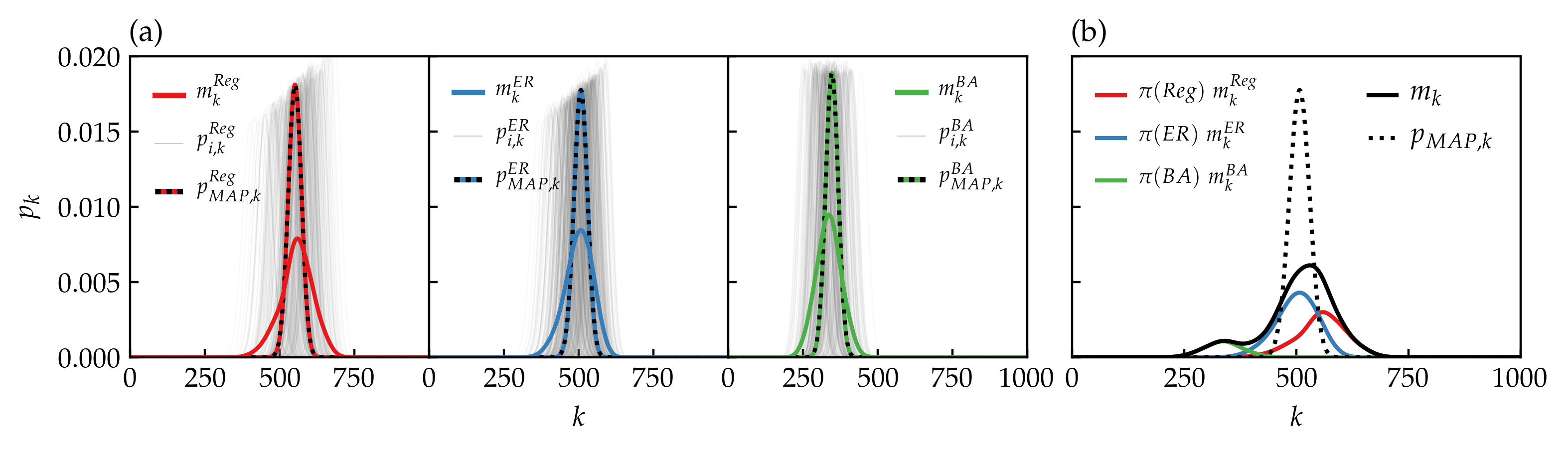}
     \caption{Illustration of the possible types of predictions for a single point in time during the growth phase of the epidemic. Panel (a) shows the conditional mean $m_k^{\theta}$ of the pushforward measures $\nu_{\theta,k}$ (Eq.~\ref{eq:mktheta}) together with the pushforward of the mode of $\pi_{\theta}$ for $\theta \in \Theta$. The grey lines indicate the pushforward of the samples drawn from $\pi_{\theta}$. Panel (b) shows the resulting predictions $m_k$ (Eqs.~\ref{eq:mk},\ref{eq:mk_sample}) and $p_{MAP,k}$ (Eq.~\ref{eq:predMAP}). The example shown here is a medium epidemic on an ER network (see Table~\ref{table:parameters}) with inference based on 10 observations $y=(k_1,\dots,k_{10})$ approximately equally spaced in time between $k_1=50$ and $k_{10}=160$}
     \label{fig:predexample}
\end{figure}

To obtain predictions one needs to integrate Eq.~\ref{eq:BD} with the parameters $u=(C,\alpha,p)$ obtained from the posterior $\pi_\theta(u|y,s)$. We generate and evaluate two different types of predictions. The first variant incorporates information on prediction uncertainty as encoded in $\pi(\theta)$ and $\pi_{\theta}(u)$. The second variant is based on a point estimate of $u$. The Python routine for generating the predictions will be made available at \url{https://github.com/tzerenner/EpidemicPredictionsFromPopulationLevelData}.

For brevity, in the following, we neglect the dependence on $t$ and instead consider some fixed point in time. We denote by $\nu_{\theta,k}$ the pushforward measure of $\pi_{\theta}$ under the forward solution operator $G_k: u \mapsto p_k$ defined by Eq.~\ref{eq:BD}. The density of $\nu_{\theta,k}$ is then related to that of $\pi_\theta$ through 
$\nu_{\theta,k}(p_k)=\pi_{\theta}(G_k^{-1}(p_k))$. The conditional mean of $\nu_{\theta,k}$ is given by

    \begin{equation}
        m_{k}^{\theta} = \int p_k~\nu_{\theta,k}(p_k)~dp_k = \int G_k(u)~\pi_\theta(u)~du.
        \label{eq:mktheta}
    \end{equation}
    
When additionally integrating over all network classes $\Theta$, we can further obtain the conditional mean of $\nu_{k}$, the pushforward measure of $\pi$, as

    \begin{equation}
        m_{k} =  \int p_k~\left(\sum_{\theta \in \Theta} \pi(\theta)~\nu_{\theta,k}(p_k)\right)~dp_k =  \int G_k(u)~\left( \sum_{\theta \in \Theta} \pi(\theta) \pi_\theta(u) \right)~du.
        \label{eq:mk}
    \end{equation}
    
We estimate $m_k$ from samples $(C,\alpha,p)_{i,1\leq i \leq n}$ which we draw from the posterior distributions $\pi_{\theta}(C,\alpha,p),~\theta \in \Theta$, using the Metropolis–Hastings algorithm. Since integrating the Master equation with a large number of parameter combinations is computationally demanding, we thin the samples by including only every $i$-th draw, such that the auto-correlation between subsequent draws is $<0.1$. To choose an appropriate size $n$ of the (thinned) sample, we consider its multivariate effective sample size (mESS), which is estimated by

\begin{equation*}
    \widehat{\mathrm{mESS}} = n \left( \frac{\mathrm{det}(\Lambda_n)}{\mathrm{det}(\Sigma_n)} \right)^{1/3},
\end{equation*}

where $\Lambda_n$ denotes the sample covariance and $\Sigma_n$ denotes the multivariate batch mean estimator of the covariance matrix in the Markov chain central limit theorem \citep{roy2020convergence, vats2020analyzing}. The sample size $n$ is chosen to be the minimum $n$ such that $\widehat{\mathrm{mESS}} \geq 260$ which ensures a confidence level of $\delta = 0.1$ and a tolerance level of $\epsilon = 0.25$ for the expectation in the three parameter $(C,\alpha,p)$-space \citep{vats2019multivariate}. To compute the mESS and the threshold for the desired confidence and tolerance levels, we used the Python implementation available at \url{https://github.com/Gabriel-p/multiESS}.

We then proceed to integrate the Master equation with each $(C,\alpha,p)$-triplet to obtain $p^{\theta}_{i,k}$ and finally approximate the conditional mean by

    \begin{equation}
        m_{k} = \sum_{\theta \in \Theta} \left( \pi(\theta)~\sum_i \frac{p^{\theta}_{i,k}}{n} \right),
        \label{eq:mk_sample}
    \end{equation}
    
as well as the respective cumulative density over $k$ by
    \begin{equation}
            M_k = \sum_{x\leq k} m_x.
        \label{eq:mk_cdf}
    \end{equation}  

From the latter we obtain equal-tailed credible intervals for the predicted number of infected nodes. Equal-tailed intervals are defined such that the probability of being below the interval is as high as being above the interval and thus can be directly obtained from the quantiles of the cumulative density as

    \begin{equation}
        Q(x) = \mathrm{inf} \{ k \in \{0, \dots, 1000\}: x \leq M_k \}.
        \label{eq:qi}
    \end{equation}

The interval $[Q(0.05),Q(0.95)]$ for example corresponds to the 90\% equal-tailed interval. When obtained from $m_k$ (Eq.~\ref{eq:mk}), such intervals incorporate both prediction uncertainty arising from uncertainty about parameters and network class as encoded in $\pi_{\theta}(u)$ and $\pi(\theta)$, respectively, as well as prediction uncertainty arising from the intrinsic stochasticity of the epidemic spreading.

The second prediction variant is based on a point estimate, i.e., a single $u=(C,\alpha,p)$ inferred from the posterior $\pi$. We first identify the most likely network class, i.e, the mode of $\pi(\theta)$,

    \begin{equation}
        \hat{\theta}_{MAP} = \mathrm{argmax}_{\theta} \{ \pi(\theta|y,s)\},
        \label{eq:maptheta}
    \end{equation}

where MAP stands for maximum a-posteriori, and then estimate the mode of $\pi_{\hat{\theta}}$,

    \begin{equation}
        \hat{u}_{\hat{\theta}, MAP} = \mathrm{argmax}_{u} \{ \pi_{\hat{\theta}}(u|y,s) \},
        \label{eq:mapu}
    \end{equation}

using a combination of global and local optimisation routines \citep{di2020network}. The predictions $p_{MAP,k}$ are obtained by integrating the Master equation with $\hat{u}_{\hat{\theta}, MAP}$, i.e.,

    \begin{equation}
        p_{MAP,k} = G_k(\hat{u}_{\hat{\theta}, MAP}).
        \label{eq:predMAP}
    \end{equation}

Again, we compute the respective cumulative density over $k$ as 
    \begin{equation}
            P_{MAP,k}  = \sum_{x\leq k} p_{MAP,x},
    \end{equation}  
    
from which we can obtain quantiles and equal-tailed prediction intervals. Such intervals are not credible intervals in the Bayesian sense, but solely represent the intrinsic stochasticity of the epidemic spreading. They are thus systematically narrower than the credible intervals discussed above. An illustration of the two prediction variants for one single point in time is provided in Fig.~\ref{fig:predexample}.

To compare uncertainty in the $p_k$-space for the two prediction variants we further consider the covariance of the pushforward $G: u \mapsto (p_0,p_1,\dots p_k,\dots p_N)^T$,  $G(u)=(G_1(u),G_2(u),\dots G_k(u),\dots G_N(u))^T \in \mathbb{R}^{(N+1)}$. We denote the mean of the pushforward of $\pi_\theta(u)$ under the forward solution operator $G$  by $m=(m_0,m_1,\dots m_k,\dots m_N)^T \in \mathbb{R}^{(N+1)}$ (Eq.~\ref{eq:mk}). Its covariance $\mathcal{C}= \left(\mathcal{C}_{kl}\right)\in  \mathbb{R}^{(N+1)\times(N+1)}$ is given by the outer product 

\begin{equation*}
  \mathcal{C} = \int (G(u)-m)(G(u)-m)^T~\pi_\theta(u)~du.
  \label{eq:C}
\end{equation*}

We estimate $\mathcal{C}$ from the discrete samples (Eq.~\ref{eq:mk_sample}) as

\begin{equation*}
  \mathcal{C}_{kl}= \sum_{\theta \in \Theta} \left( \frac{1}{n-1} \sum_{i=1}^n (p_{k,i}-m_k)(p_{l,i}-m_l)~\pi(\theta) \right),~~k=0,\dots N,~l=0\dots N. 
\end{equation*}

We can then evaluate 

\begin{equation}
    |\mathcal{C}|  = \left| \int (G(u)-m) (G(u)-m)^T~\pi_\theta(u)~du \right|,
    \label{eq:normC}
\end{equation}

where $|\cdot|$ denotes the Euclidean norm in $\mathbb{R}^{(N+1)\times(N+1)}$. To accordingly evaluate the uncertainty of the predictions based on the point estimator from Eq.~\ref{eq:mapu}, we further evaluate 

\begin{equation}
    |\mathcal{C}_{MAP}|  = \left| \int (G(u)-p_{MAP}) (G(u)-p_{MAP})^T~\pi_\theta(u)~du \right|,
    \label{eq:normCMAP}
\end{equation}

where $p_{MAP}=(p_{MAP,0},p_{MAP,1},\dots m_{MAP,k},\dots m_{MAP,N})^T \in \mathbb{R}^{(N+1)}$ are the predictions from Eq.~\ref{eq:predMAP}.

        \subsection{Performance assessment}

We generate a reference for evaluating the BD model by carrying out a set of Gillespie simulations which provides us with an i.i.d. sample of the stochastic spreading of an SIS epidemic on a given network. We chose a sample size of 1000 and denote the set of 1000 epidemic trajectories obtained from the sample as $\{I_{r,i}(t)\}_{1\leq i \leq 1000}$. To empirically validate the BD model, we compare the $p_k(t)$ obtained from the numerical integration of Eq.~\ref{eq:BD} against the reference. We evaluate the difference in expectation, i.e.,

\begin{equation}
    \Delta \hat{I}(t) = \widehat{I}_{s}(t) - \widehat{I}_{r}(t) =  \sum_{k=0}^{N} k~p_{k}(t) - \sum_{i=1}^{1000} \frac{I_{r,i}(t)}{1000},
    \label{eq:deltaI}
\end{equation}

where $\widehat{I}_{r}(t)$ denotes the mean over the reference at time $t$ and $\widehat{I}_{s}(t)$ denotes the mean number of infected nodes at time $t$ predicted by the BD model. We further evaluate the cumulative densities from the BD model in comparison to our reference using the integrated quadratic distance (IQD) which is given by

\begin{equation}
    \mathrm{IQD}(t)=\int_{-\infty}^{\infty}\left(F_{s,t}(x)-F_{r,t}(x)\right)^2 dx.
    \label{eq:iqd}
\end{equation}

Here $F_{s,t}(x)$ denotes the cumulative density at time $t$ predicted by the BD model,

\begin{equation*}
   F_{s,t}(x) = \sum_{k\leq x} p_{k}(t),
\end{equation*}

and $F_{r,t}(x)$ denotes the reference empirical cumulative density,

\begin{equation*}
   F_{e,t}(x) = \frac{1}{1000} \sum_{i=1}^{1000}  \mathbf{1}_{I_{r,i}(t) \leq x},
   \label{eq:ecdf}
\end{equation*}

where $\mathbf{1}$ is an indicator function equal to $1$ if condition $I_{r,i}(t) \leq x$ is true and 0 otherwise. A small IQD is obtained when not only the mean but also the intrinsic stochasticity of the epidemic spreading is adequately represented by the BD model. The goodness of fit between the cumulative densities is further illustrated by quantile-quantile plots. The quantile function is the inverse of the cumulative density, i.e., $Q_{s/r,t} = F^{-1}_{s/r,t}$, and hence given by

\begin{equation}
    Q_{s/r,t}(P) = \mathrm{inf}\{x \in \{1, \dots,1000\}: P \leq F_{s/r,t}(x) \},
    \label{eq:quantiles}
\end{equation}

for the BD model (s), and the reference (r), respectively.
        
    \section{Data}
        \label{sec:data}

        We consider nine different network and epidemic parameter combinations. The network parameters, network class and mean degree $\langle k \rangle$, and the epidemic parameters, per link infection rate $\tau$ and recovery rate $\gamma$, are summarised in Table~\ref{table:parameters}. Gillespie simulations were performed on a network of $N=1000$ nodes and initialised with five infected nodes selected at random. The time $T$ is an approximate value of the time span between initialising the simulation and reaching quasi-steady state and in the following serves as a universal time scale which allows to plot the different cases onto the same time axis. Time is unit-free here. The simulated data can be re-scaled to physically-meaningful time scales by applying an appropriate multiplicative factor to the simulation time $t$. The parameters for the nine cases were chosen such that we obtained one large epidemic with $>70\%$ of the population infected at quasi-steady state, one medium epidemic with $40\%$ to $60\%$ of the population infected at quasi-steady state, and one small epidemic with $<40\%$ of the population infected at quasi-steady state for each network class.

        In this study, we consider the epidemics at population-level, that is, we aim to predict the future course of the number of infected nodes. Network class and parameters of the BD model are inferred from population-level observations of the number of infected nodes at a set of discrete points in time. 
        
        \begin{table}[h]
        \centering
        \begin{tabular}{llllll}
        \hline
        case  & $\langle k \rangle$  & $\tau$ & $\gamma$ &  $T$ \\
        \hline
        Reg l & 5	& 4.251 & 2.969 & 0.75 \\
        Reg m & 10 & 1.265 & 5.773 & 1.5 \\
        Reg s & 7	& 0.762	& 3.356 & 8 \\
        \hline
        ER l & 8.124 & 1.251 & 0.969 & 1.25 \\
        ER m & 15.868 & 0.859 & 6.338 & 1.25 \\
        ER s & 12.042 & 1.143 & 9.579 & 2 \\
        \hline 
        BA l & 13.902 & 3.123 & 6.969 & 0.25 \\
        BA m & 9.95 & 2.19 & 8.948 &	0.5 \\
        BA s & 7.968 & 0.612 & 3.803 & 2.5 \\
        \hline
        \end{tabular}
        \caption{Parameters of the simulated SIS epidemics on networks of $N=1000$ nodes. Listed are the names of the different cases, which consist of the respective network class (Regular (Reg), Erd\H{o}s–R\'{e}nyi (ER) or Barab\'asi-Albert (BA)) and epidemic size (large (l), medium (m), small (s)), mean node degree $\langle k \rangle$, per-link infection rate $\tau$, recovery rate $\gamma$ and the approximate time $T$ between initialisation and quasi-steady state in simulations initialised with five infected nodes selected at random.}
        \label{table:parameters}
        \end{table}

    \section{Results}
        
  \subsection{Validation of the BD model}
  \label{sec:results1}

    We carry out a set Gillespie simulations during which we keep track of the number of infected nodes $k$, the number of S-I links over time and the time spent in the observed states. For each case, we carry out in total 200 simulations half of which are initialised with five infected nodes and half with 1000 infected nodes. With this choice of initial conditions, we obtain realisations of the random variable $a_k$ ($\tau~\times$ \#S-I links) for each $k=0,\dots,N$, from which we compute the expectations $\hat{a}_k$ following \cite{di2020network} as 
        
        \begin{equation}
            \hat{a}_k = \tau \frac{\sum_i i~t_{ik}}{\sum_i t_{ik}},~~~k=1,\dots,N,
            \label{eq:ak_hat}
        \end{equation}
        
        where $t_{ik}$ denotes the total lifetime of all network states with $k$ infected nodes and $i$ S-I links.
        
        \begin{table}[t]
        \centering
        \begin{tabular}{lcccr}
        \hline
        case & $C \times 10^4$ & $\alpha$ & $p$ & RMSE \\
        \hline
        Reg l & 0.469  & 0.475  & 0.915 & 85.10\\
        Reg m & 0.110  & 0.182  & 1.007 & 6.90\\
        Reg s & 0.039  & 0.224  & 1.019 & 6.41\\
        \hline
        ER l & 0.116  & -0.085 &  0.977 & 36.48\\
        ER m & 0.162  & 0.020   & 0.984 & 18.07\\
        ER s & 0.169  & 0.016  & 0.983 & 20.01\\
        \hline
        BA l & 4.872  & -0.726 & 0.799 & 208.09\\
        BA m & 3.229  & -0.551 & 0.778 & 241.24\\
        BA s & 0.765  & -0.494 & 0.776 & 56.07\\
        \hline 
        \end{tabular}
        \caption{$(C,\alpha,p)$-triples from a least-squares fit of the parametric $a_k$ model (Eq.~\ref{eq:cap}) to $\hat{a}_k$ from Gillespie simulations (Eq.~\ref{eq:ak_hat}) for all nine cases (see~Table \ref{table:parameters}). The right column shows the root mean square error between empirical $\hat{a}_k$ and parametric model $\mathrm{RMSE}(\hat{a}_k,a_k(C,\alpha,p))$.}
        \label{table:cap}
        \end{table}

    We then proceed to fit the parameters $(C,\alpha,p)$ of the parametric $a_k$ model from Eq.~\ref{eq:cap} to the $(k,\hat{a}_k)$ curves by a least-squares fit using a particle swarm algorithm. Table~\ref{table:cap} lists the resulting $(C,\alpha,p)$-triples for each case along with the root mean square error (RMSE) between parametric $a_k$ curves and $\hat{a}_k$. 
        
     Figure~\ref{fig:ak_curves} shows the empirical $\hat{a}_k$ curves as well as the fitted $a_k(C,\alpha,p)$ curves for the nine cases. The top panels illustrate the good agreement between the parametric model and $\hat{a}_k$. The bottom panels of Fig.~\ref{fig:ak_curves} show relative errors. It is not surprising that the relative error is largest for $k$ close to zero or close to $N$, i.e., when the number of S-I links is small either because only very few nodes are infected or because almost the entire population is infected. The relative errors are lowest for the ER network class followed by the Regular networks. For the BA network class, we obtain larger errors.
 
        \begin{figure}[t]
          \centering
          \includegraphics[width=\linewidth]{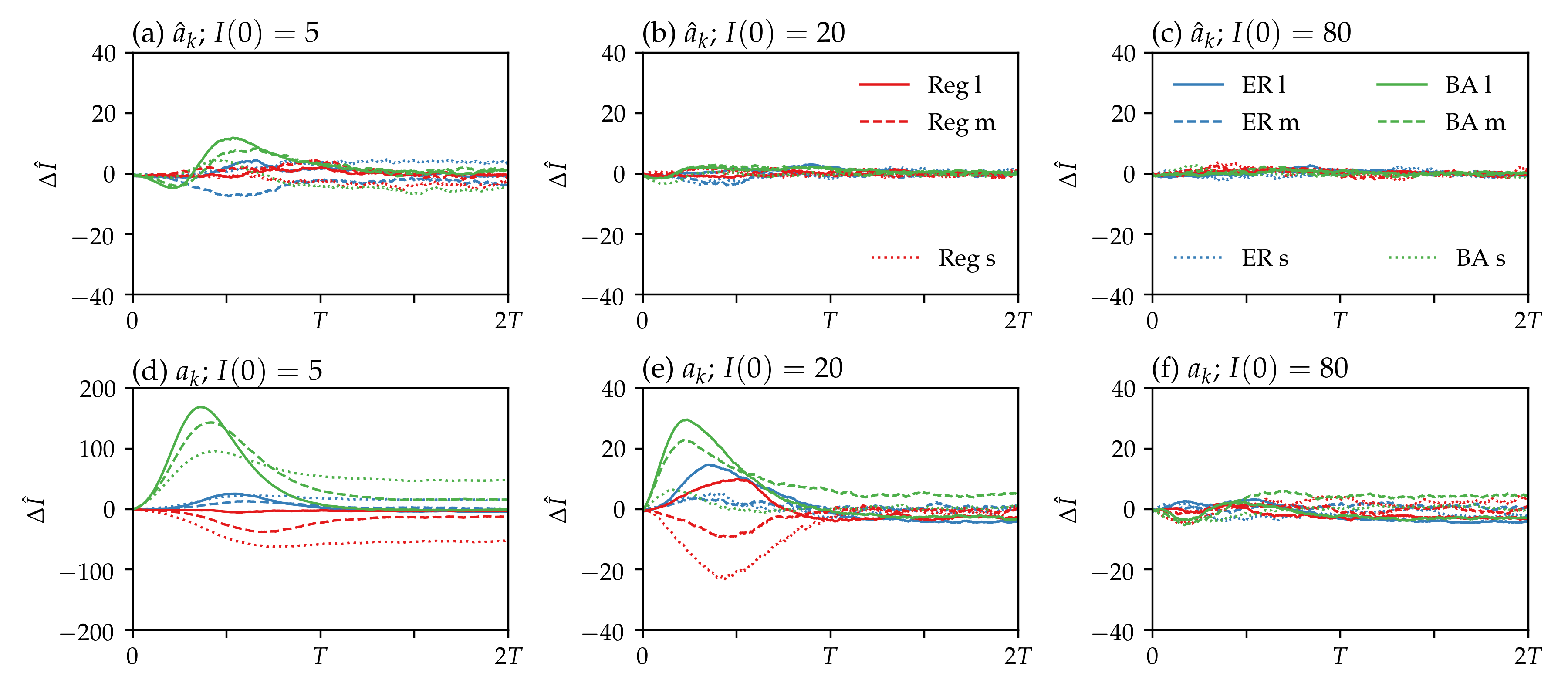}
          \caption{Difference in expected number of infected nodes between BD model and reference vs. time (Eq.~\ref{eq:deltaI}). The top panels show the BD model with the empirical $\hat{a}_k$. The bottom panels show the BD model with parametric $a_k(C,\alpha,p)$. From left to right, three different initial conditions are shown: $I(t_0)=5,20,80$ at $t_0=0$. The different colours and line styles indicate the nine different cases (see~Table~\ref{table:parameters}).}
          \label{fig:bias}
         \end{figure}
        
        \begin{figure}[t]
          \centering
          \includegraphics[width=\linewidth]{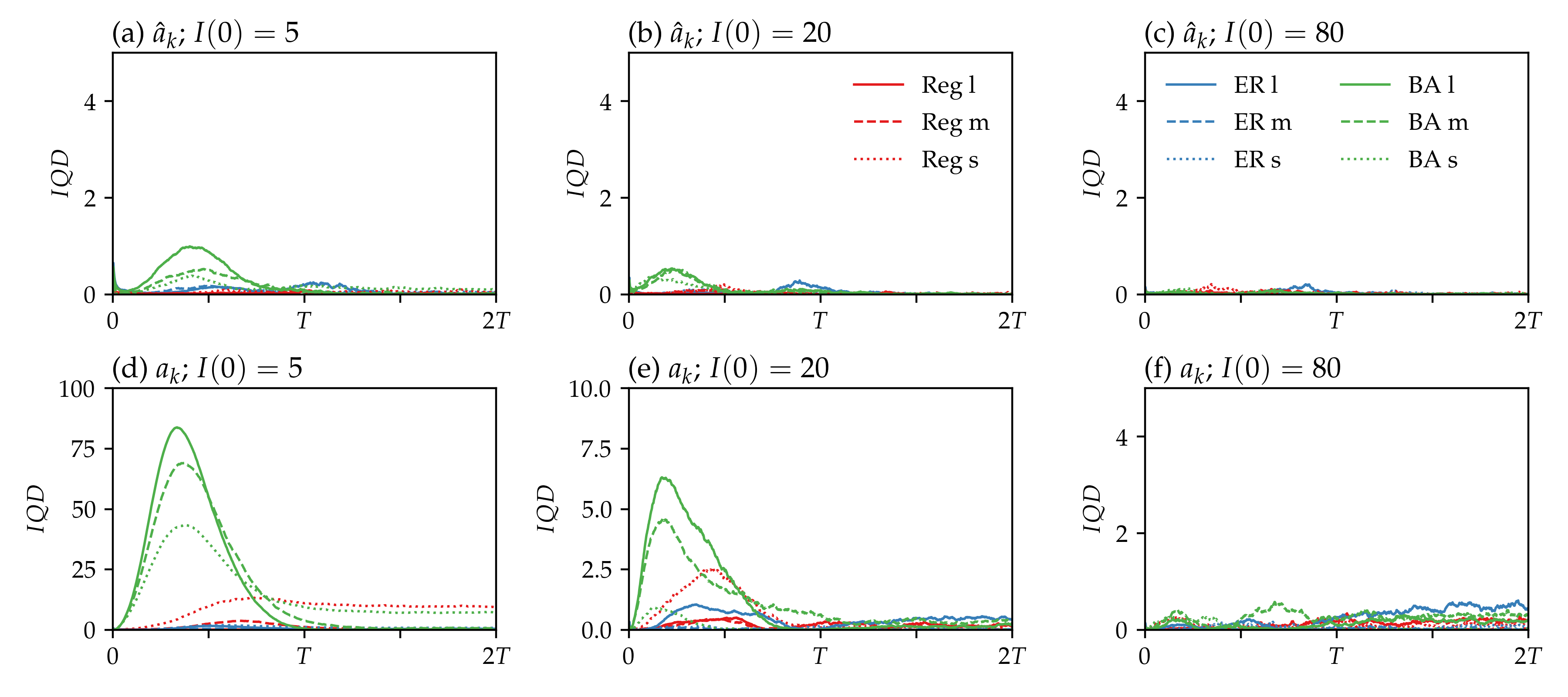}
          \caption{Integrated quadratic distance between between the cumulative density from the BD model and the reference vs. time (Eq.~\ref{eq:iqd}). The top panels show the BD model with the empirical $\hat{a}_k$. The bottom panels show the BD model with parametric $a_k(C,\alpha,p)$. From left to right, three different initial conditions are shown: $I(t_0)=5,20,80$ at $t_0=0$. The different colours and line styles indicate the nine different cases (see~Table~\ref{table:parameters}).}
          \label{fig:iqd}
         \end{figure}

        \begin{figure}[t]
          \centering
          \includegraphics[width=0.9\linewidth]{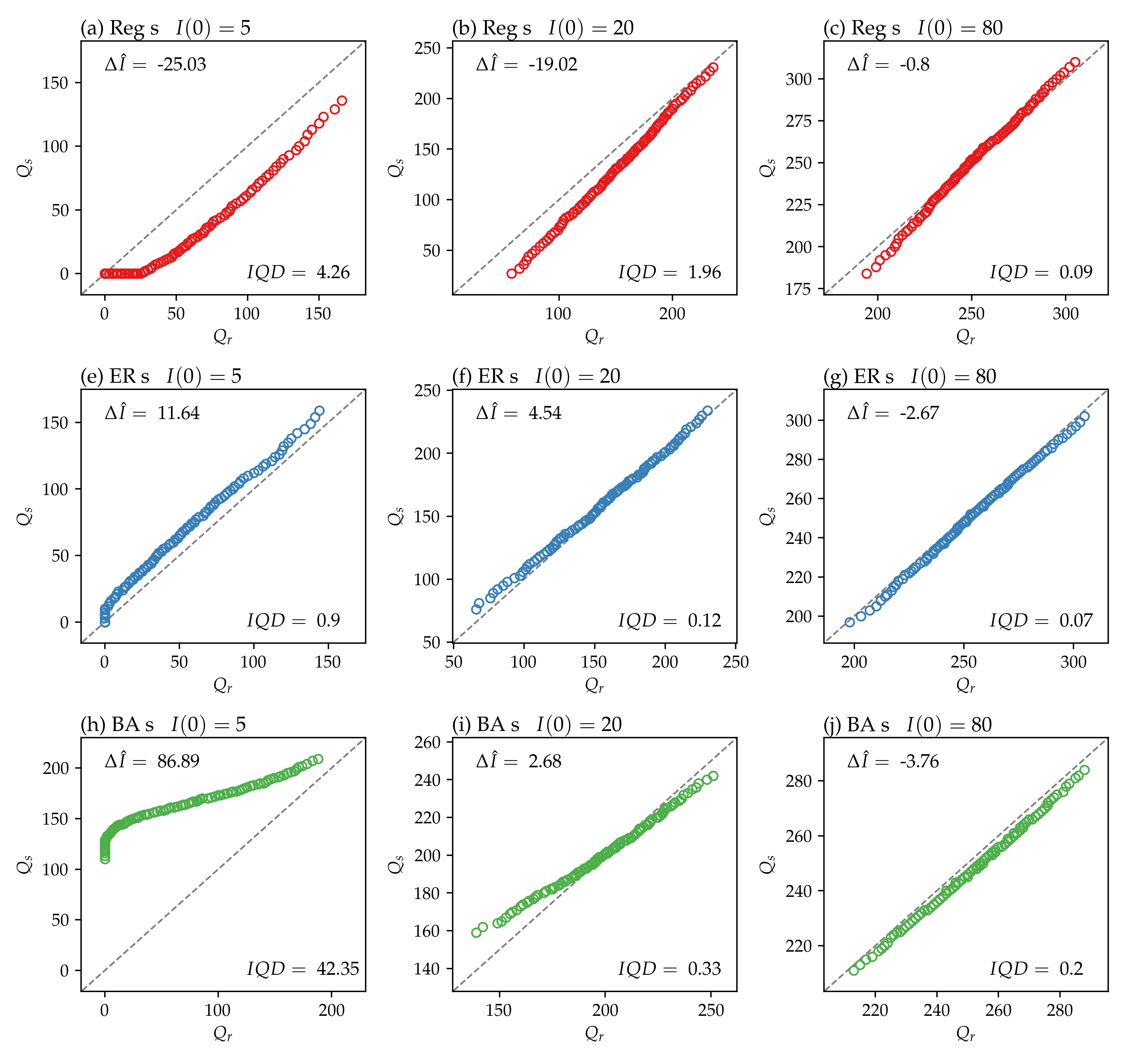}
          \caption{Quantile-quantile plots comparing BD model predictions (s) and reference (r). The circles indicate the quantiles $Q(0.05)$, $Q(0.06), \dots Q(0.95)$ at time $t=t_0+\frac{1}{3}T$ after initialisation with $I(t_0)=5,20,80$ infected nodes at time $t_0=0$.}
          \label{fig:qqplot}
        \end{figure}

    We simulate each case by integrating the Master equation (Eq.~\ref{eq:BD}) with the empirical $\hat{a}_k$ as well as the parametric $a_k(C,\alpha,p)$. We initialise simulations with $I(t_0)=5,~20$ and 80 infected nodes at time $t_0=0$, the latter two values corresponding to two and four cycles of doubling from the initially five infected nodes. A set of 1000 Gillespie simulations of each case serves as a reference. Figure~\ref{fig:bias} shows the difference in the expected number of infected nodes between BD model and reference. Figure~\ref{fig:iqd} shows the integrated quadratic distance (IQD) between the cumulative densities from BD model and reference. The errors remain small throughout the simulations with the empirical $\hat{a}_k$ (top panels), which confirms the suitability of the BD model to describe population-level infection rates in SIS epidemics on Reg, ER and BA networks. Not only is the expectation well captured by simulations with $\hat{a}_k$, but also the intrinsic stochasticity of the epidemic despite the mean-field approximation. We observe the largest errors for the BA network class and during the growth phase when simulations are initialised with $I(t_0)=5$ (Fig.~\ref{fig:iqd}a). The BA network class exhibits a higher degree of heterogeneity than Regular and ER networks. Hence, a larger variance in the number of S-I links at a given $k$ is expected. Therefore, the mean-field approximation might be less well suited for that type of network than for Regular and ER networks.
        
    The simulations with the BD model with the parametric $a_k(C,\alpha,p)$ (bottom panels) exhibit larger errors compared to the simulations with $\hat{a}_k$. Ranking the different cases studied, the BD model achieves the lowest errors for the ER network class, followed by Regular networks and the BA networks. Again, errors are largest when the simulations are initialised with $I(0)=5$ which appears to be caused by the larger relative errors of the parametric $a_k$-model for small $k$. The overestimation of $a_k$ at small $k$ by the parametric model for the BA network class (Fig.~\ref{fig:ak_curves}c) causes the number of infected nodes to increase too fast in the BD simulations, which leads to an over-estimation of the number of infected nodes during the growth phase (Figs.~\ref{fig:bias}d,~\ref{fig:qqplot}h). Conversely, the underestimation of $a_k$ for small $k$ for the Regular networks (Fig.~\ref{fig:ak_curves}b) causes the number of infected nodes to increase too slowly in the BD model simulations, and hence an under-estimation of the number of infected nodes during the growth phase (Figs.~\ref{fig:bias}d,~\ref{fig:qqplot}a). The errors peak during the growth phase and then decay until reaching an approximately constant value in the quasi-steady state. 
        
    For the majority of the cases, the quasi-steady state is well captured in both mean and variation around the mean, with the only exception being the small epidemics on Regular and BA networks. When the BD model is initialised at $I(0)=5$, it starts from with a state from which some of the small epidemics will eventually die out and only some will eventually converge to the quasi-steady state. Due to the over-estimation of $a_k$ for small $k$ for the BA networks in the parametric model, the probability of an epidemic to proceed to the quasi-steady state from $I(0)=5$ is over-estimated in the BD model. Hence, the expected number of infected nodes in the BD model is too large. For Regular networks, the opposite holds and the expected number of infected nodes is too small. When initialised with $I(0)=20$ the errors are smaller, but the temporal pattern of the errors persists, i.e., errors peak during the growth phase and then decay until the quasi-steady state is reached (Figs.~\ref{fig:bias}e,~\ref{fig:iqd}e). When initialised with $I(0)=80$, we find that both the growth phase as well as the quasi-steady state is well captured by the BD model (Figs.~\ref{fig:bias}f,~\ref{fig:iqd}f,~\ref{fig:qqplot}c,g,j).
        \subsection{Predictions}
\label{sec:results2}

\subsubsection{Network class inference}

\cite{di2020network} demonstrated that one can reliably recover the class of the underlying network from population-level observations, when observations of the full epidemic trajectory from an early stage up to quasi-steady state are available. When aiming to predict the future evolution of an ongoing epidemic, the inference of network class and parameters $(C,\alpha,p)$ can only utilise observations of the epidemic up to its current state. Therefore, the question arises as to when one has sufficient information during an epidemic to reliably predict its further course. We therefore carry out a sensitivity analysis by inferring the posterior distribution $\pi(\theta),~\theta \in \Theta=\{\rm{Reg, ER, BA}\}$ from observation data sets covering different time windows during the evolution of the epidemic and incorporating different numbers of observations. For this analysis, we consider the medium epidemics (Table~\ref{table:parameters}). 

    \begin{figure}[h!]
    \centering
    \includegraphics[width=\linewidth]{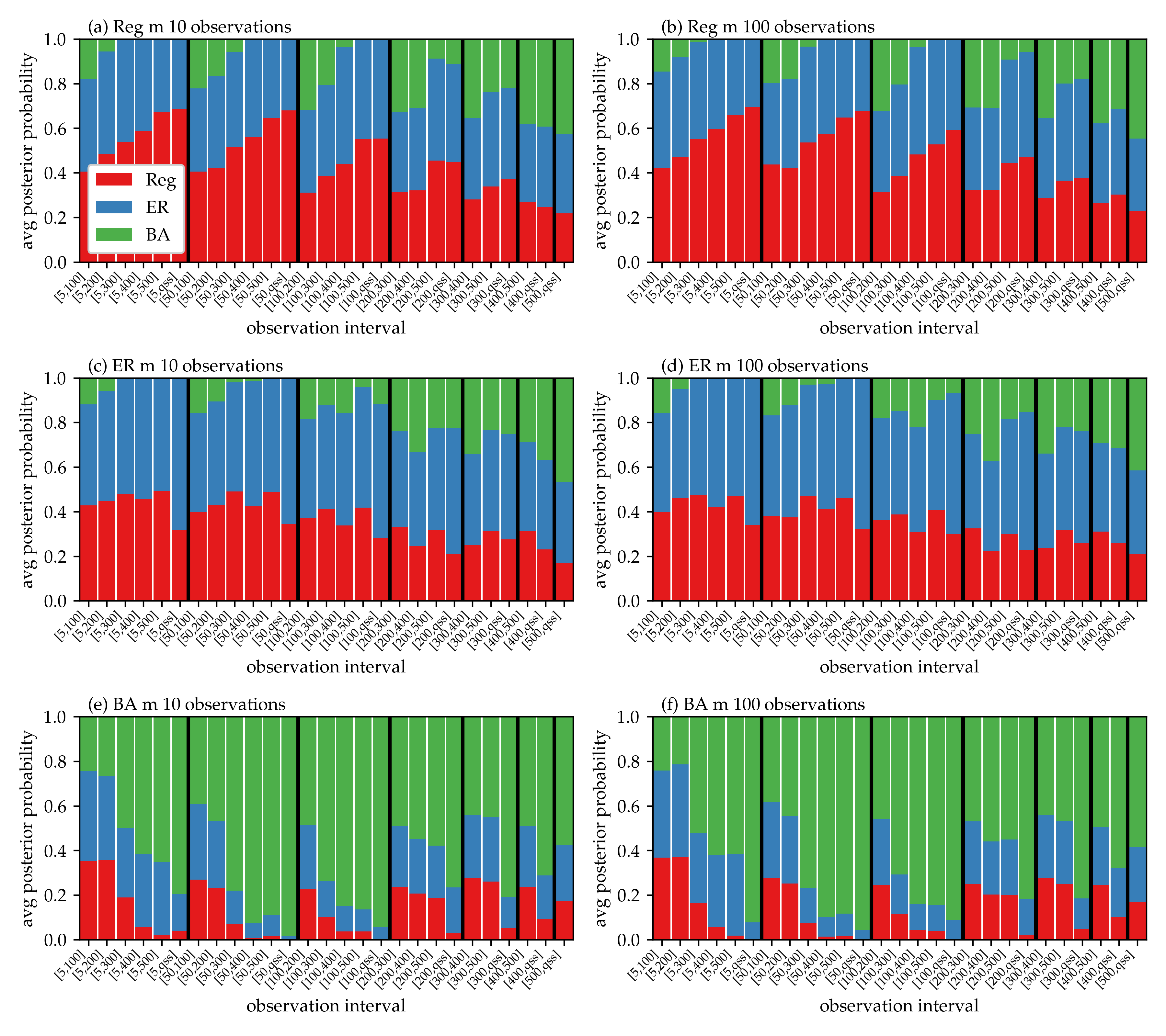}
    \caption{Sensitivity of network class inference on observational time span and number of observations. The bars indicate the average posterior probability $\pi(\theta),~\theta \in \Theta=\{\rm{Reg, ER, BA}\}$ over ten realisations of the medium epidemic on Regular, ER or BA network (Table~\ref{table:parameters}). For each case, 27 different observation intervals have been evaluated. The ten (or 100) observations are spaced approximately equidistantly in time throughout the observational period.}
    \label{fig:ClassSens}
    \end{figure}

    \begin{figure}[t]
    \centering
    \includegraphics[width=\linewidth]{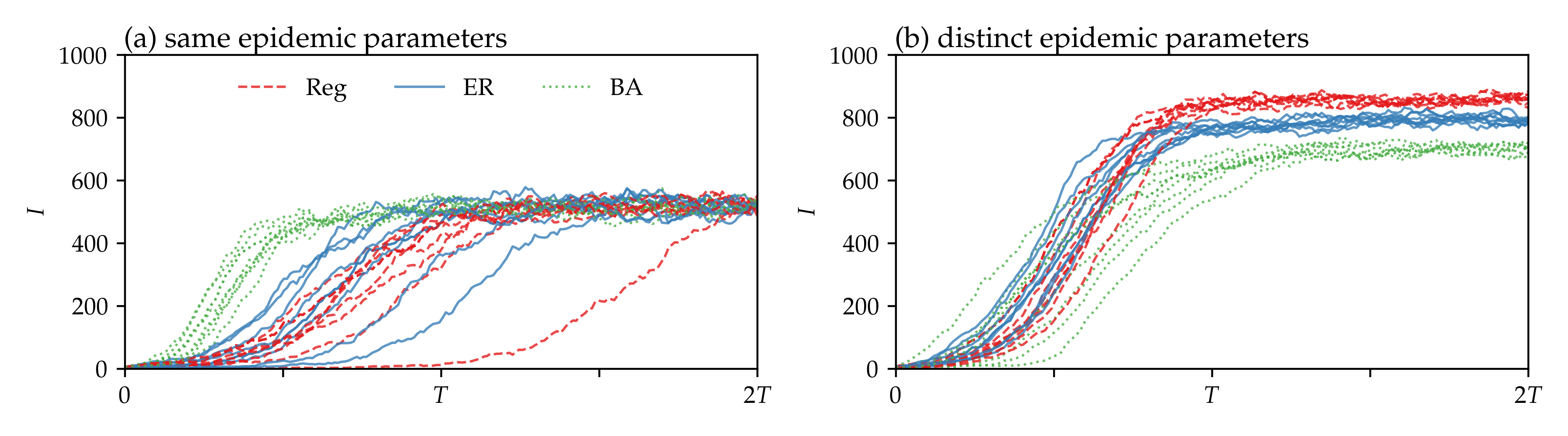}
    \caption{Examples of epidemic trajectories from Gillespie simulations on Regular, ER and BA networks. In panel (a), the epidemic parameters $\tau = 1.265,~\gamma = 5.773 $ and average node degree $k\approx 10$ (ER m, Table~\ref{table:parameters}) are (approximately) the same for all three networks. In panel (b), epidemic parameters and average node degree are distinct for the three different networks and chosen such that trajectories exhibit a similar course during the early stage of the epidemic: $\tau_{ER} = 3.5,~\gamma_{ER} =2.969,~ k_{ER}= 5.046$;~$\tau_{Reg} = 4.251,~\gamma_{Reg} = 2.969,~k_{Reg} = 5$ (Reg l, Table~\ref{table:parameters});~$\tau_{BA} = 3.2,~\gamma_{BA} = 2.969,~k_{BA} = 3.992$. Eight realisations are shown for each network class and parameter combination.}
    \label{fig:trajcetories}
    \end{figure}

The results are summarised in Fig.~\ref{fig:ClassSens}. As expected, in general, the longer the observational period, the higher the (average) posterior probability of the true underlying network class is. When the observational period ranges from an early stage of the epidemic up to quasi-steady state (50 to quasi-steady state), ten out of ten realisations on the BA network and seven out of ten realisations on Regular and ER networks are classified correctly. While BA networks can be clearly separated when sufficient data is available, distinguishing between Regular and ER networks appears challenging. For some realisations, the respective trajectories largely overlap  (Fig.~\ref{fig:trajcetories}a). 

When the observation time span is shortened, the rate of correct classifications decreases (Fig.~\ref{fig:ClassSens}). As demonstrated in Fig.~\ref{fig:trajcetories}b, epidemics spreading on networks from the different classes may exhibit a similar shape during the earlier stage and only diverge later on. Thus, inferring the underlying network classes from population-level observations of a single realisation requires a sufficient observational time span. Increasing the number of observations from 10 to 100 does not have any visible effect on the classification. Ten observations provide a sufficient description of the epidemic trajectory. 

We note that for the BA networks, classification accuracy increases when the very early stage of the epidemic up to $I \approx 50$ is excluded from the observational data set. In Fig.~\ref{fig:ClassSens}e,f, this is most obvious when comparing the posterior probabilities obtained for the observation intervals $[5,400]$, $[5,500]$ and $[5,qss]$ with those obtained for $[50,400]$, $[50,500]$ and $[50,qss]$, respectively. We believe this to be caused by the relatively large error of the parametric $a_k$-model for small $k$ for the BA network class (Fig.~\ref{fig:ak_curves}c). For small $k$, the average number of S-I links and hence $a_k$ are over-estimated by the model. Hence, the initial spreading of the epidemic on the network is expected to proceed significantly slower than the BD model with optimal $(C,\alpha,p)$-triplet would suggest. Further, we note that because Regular and ER networks are comparably close in the $(C,\alpha,p)$-space, confusion between Regular and ER networks is comparably likely, but predictions are also expected to be comparably robust to confusion between Regular and ER network classes.

\subsubsection{Epidemic trajectories}

    \begin{figure}[t]
    \centering
    \includegraphics[width=\linewidth]{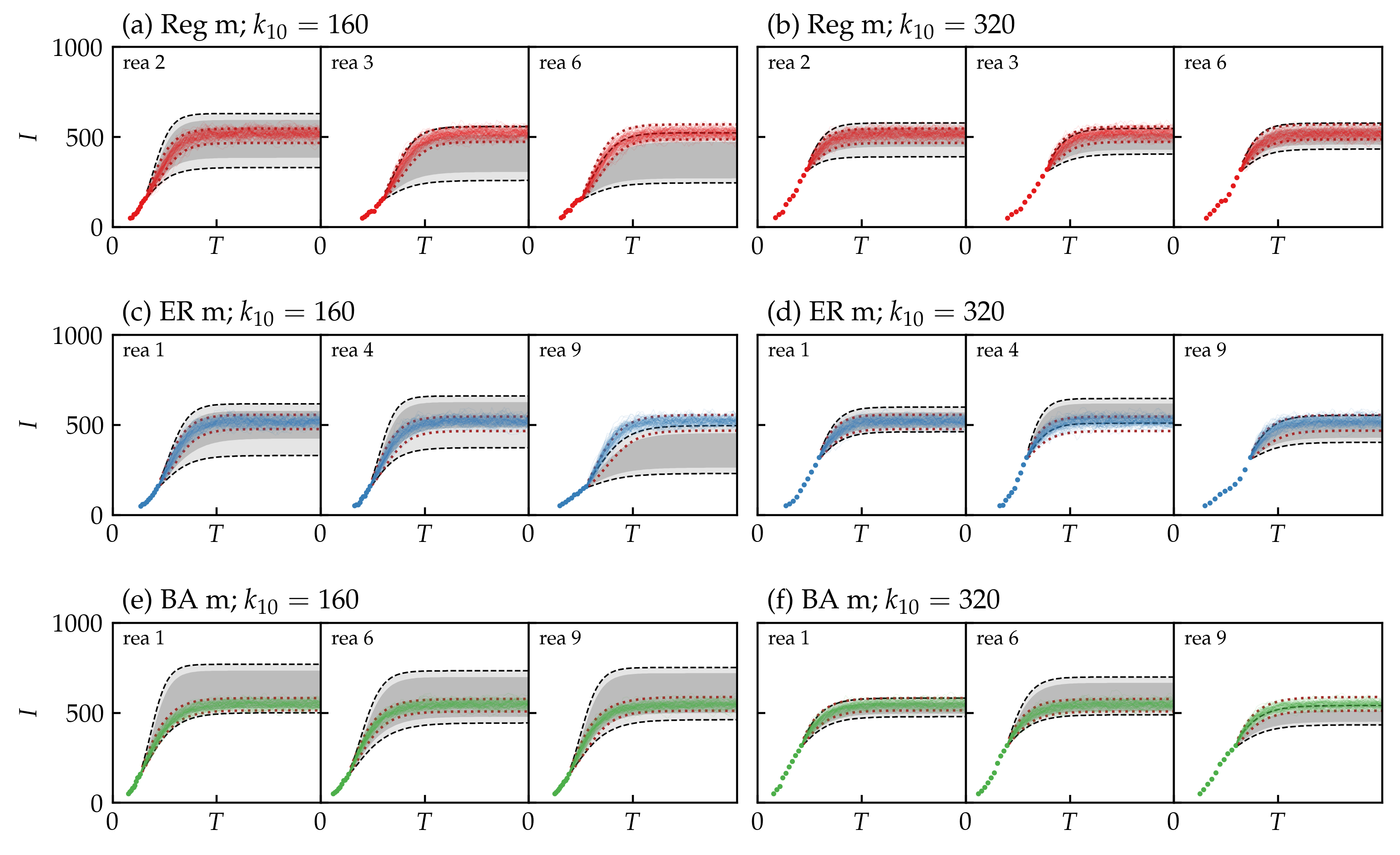}
    \caption{Predictions incorporating uncertainty for three example realisations of the medium epidemics on Regular, ER and BA networks. The grey shaded areas indicate 70\%- and 90\%-equal tailed credible intervals of the predictions initialised at the last observation $I(t_{10})=k_{10}$. The dots indicate the ten observations $(y,s)$ used for inference where $y=(k_1\approx 50 k_{10})$. The coloured lines show 100 realisations of Gillespie simulations initialised at the last observation. The dotted brown lines indicate the 90\%-equal tailed credible intervals for predictions with inference from 10 observation up to (and including) quasi-steady state.}
    \label{fig:PredUncer}
    \end{figure}

    \begin{figure}[t]
    \centering
    \includegraphics[width=\linewidth]{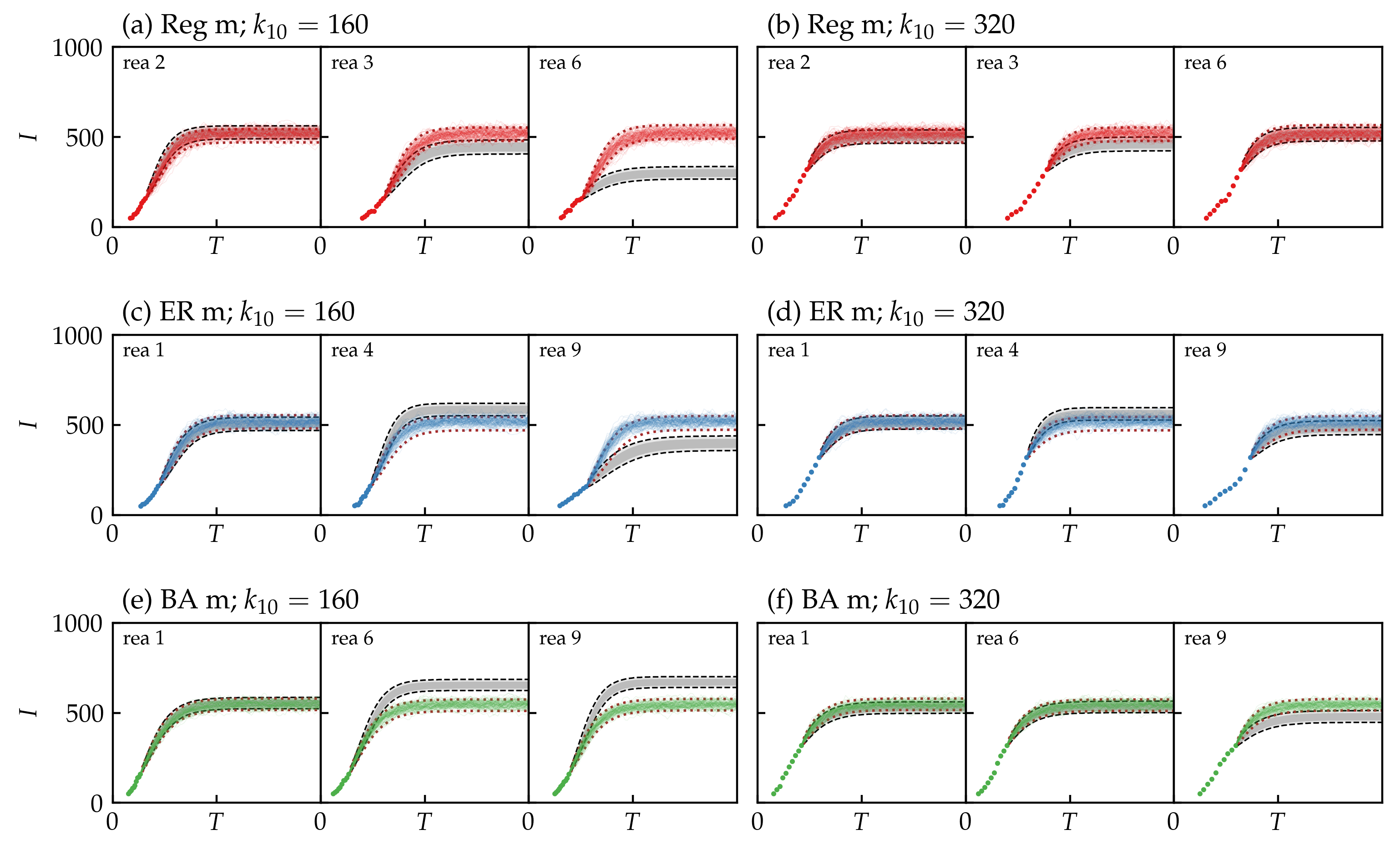}
    \caption{Point estimate-based predictions for three example realisations of medium epidemics on Regular, ER and BA networks. The grey shaded areas indicate 70\%- and 90\%-equal tailed prediction intervals of the predictions initialised at the last observation $I(t_{10})=k_{10}$. The dots indicate the ten observations $(y,s)$ used for inference where $y=(k_1\approx 50 k_{10})$. The coloured lines show 100 realisations of Gillespie simulations initialised at the last observation. The dotted brown lines indicate the 90\%-equal tailed credible intervals for predictions with inference from 10 observation up to (and including) quasi-steady state.}
    \label{fig:PredMAP}
    \end{figure}
    
    \begin{figure}[t]
    \centering
    \includegraphics[width=\linewidth]{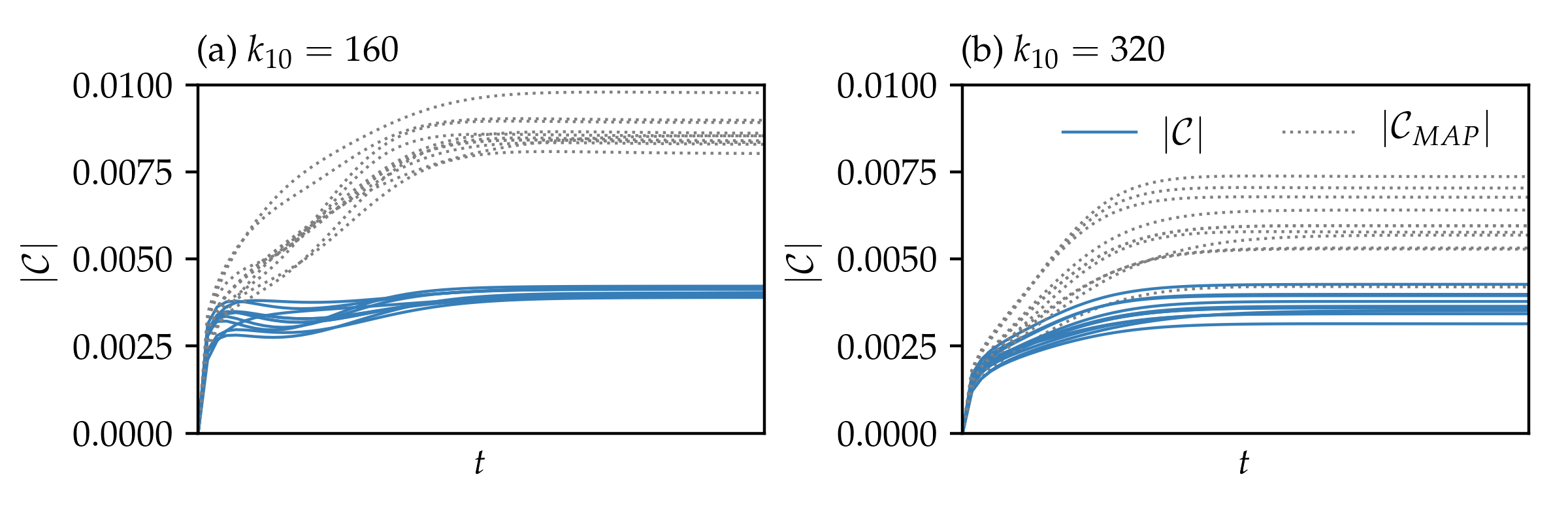}
    \caption{Uncertainty in the $p_k$-space. The solid blue lines show the Euclidean norm of the covariance $\mathcal{C}$ of the pushforward over time $t$ (Eq.~\ref{eq:normC}). The grey dotted lines show the norm of the covariance around $p_{k,MAP}$ (Eq.~\ref{eq:normCMAP}). Shown are all ten realisations of the medium epidemic on an ER network with predictions initialised at and inferred from observations up to $I(t_{10})=k_{10}$.}
    \label{fig:C}
    \end{figure}

Figure~\ref{fig:PredUncer} shows the predictions incorporating the uncertainty encoded in the posterior distribution(s). Shown are three realisations of the medium epidemics on Regular, ER and BA networks. Additional Figures for all ten realisations of the nine cases from Table~\ref{table:parameters} are provided in the Supplementary Material. The dots indicate the observations $(y,s)$. The grey-shaded areas indicate the predictions. Trajectories of 100 realisations of Gillespie simulations initialised at the network state associated with the last observational data point serve as reference.

For the majority of cases and realisations, the 90\%-credible interval (CI) contains the reference. For some cases/realisations, the reference lies just outside 90\%-CI (e.g., Fig.~\ref{fig:PredUncer}a realisation 9). For predictions of the medium epidemics initialised after five cycles of doubling ($k_{10}=160$), we find the reference to lie just outside the 90\%-CI for one out of ten realisations for the ER network, while for the Reg and BA networks, it lies partly outside of the 90\%-CI for 3 out of 10 realisations each. The 90\%-CI spans a range of up to $\approx 300$ and tends to be larger for the small and medium epidemics than for the large epidemics.

When parameters and network class are inferred from observations up to five cycles of doubling ($k_{10}=160$), prediction uncertainty is dominated by uncertainty about model parameters and network class. When inference is based on observations up to and including the quasi-steady state, uncertainty on parameters and network class is negligible and the uncertainty on the future course of the epidemic is dominated by the intrinsic stochasticity of the process (dotted brown lines in Fig.~\ref{fig:PredUncer}). Hence, the magnitude of the prediction uncertainty is sensitive to the observational time span available for inference. In any realistic setting, observations are of course not available beyond the point from which one aims to predict the future course of an epidemic. As illustrated in Fig.~\ref{fig:PredUncer}b,d and f, parameter and network class uncertainty is, as expected, reduced when a longer observational time span is available. For predictions initialised at 6 cycles of doubling ($k_{10}=320$), prediction uncertainty due to intrinsic stochasticity and parameter/network class uncertainty is similar in magnitude and it depends on the particular case and realisation if prediction uncertainty is dominated by either one. 

For the predictions of epidemics on BA networks, the reference tends to lie in the lower half of the 90\%-CI whereas for Regular networks, it tends to lie in the upper half of the 90\%-CI. Relatively large credible intervals are associated with relatively large uncertainty about the network class. As illustrated in Fig.~\ref{fig:trajcetories}b, epidemics on different networks with a similar trajectory during the early stage eventually diverge, with the epidemics on the BA networks converging to the lowest level of infection during quasi-steady state followed by ER and Regular networks. Thus, credible intervals obtained from observations of the beginning of one of the BA trajectories from Fig.~\ref{fig:trajcetories}b are expected to contain the true BA trajectories at their lower end. This behaviour can also be understood from the corresponding $a_k$-curves. Curves for networks from different classes that are similar for small $k$ will diverge for larger $k$, with the curve corresponding to the BA network having the smallest peak, typically followed by ER and finally the Regular network with the highest peak. 

Figure~\ref{fig:PredMAP} shows the point estimate-based predictions. The prediction intervals here do not account for network class and parameter uncertainty, but only represent the intrinsic stochasticity of the epidemic spreading. Accordingly, the prediction intervals are systematically narrower than the credible intervals. The width of the prediction intervals is consistent with the spread of the trajectories from the Gillespie simulations. For some cases and realisations, the point estimate-based predictions provide a near perfect fit to the reference (e.g., Fig.~\ref{fig:PredMAP}a realisation 1, e realisation 1). However for some cases/realisations prediction and reference differ by up to $\approx 300$ (e.g., Fig.~\ref{fig:PredMAP}c realisation 6). For epidemics on BA networks, the number of infected nodes is over-estimated in the point estimate-based predictions if the network is falsely identified as ER or Regular network. For epidemics on Regular networks, the number of infected nodes is under-estimated if the network is falsely identified as ER or BA. The reason for this is the same as for the tendencies of the reference to occur in different parts of the credible intervals for the different network classes discussed in the above paragraph. When inference is based on observations up to six cycles of doubling ($k_{10}=320$), the errors of the point estimate-based prediction are visibly reduced (see also the Supplementary Material). Hence, when longer observation time spans are available also point estimate-based predictions are potentially useful.

Finally, in Fig.~\ref{fig:C} we consider the uncertainty in the $p_k$-space as described by the covariance of the pushforward measure around the two different predictions $m_k$ and $p_{k,MAP}$. Shown is the medium epidemic on the ER network. As the predictions based on the conditional mean $m_k$ incorporate the uncertainty about the predicted $I(t)$ that stems from uncertainty about network class and model parameters it is systematically wider (Fig.~\ref{fig:predexample}). This width reflects the width of the pushforward and thus leads to lower values of $|\mathcal{C}|$ compared to the point-estimate based predictions.
Further, the predictions incorporating uncertainty exhibit less variation of $|\mathcal{C}|$ among the different realisations than the point estimate-based predictions. As already discussed alongside Figs.~\ref{fig:PredUncer} and~\ref{fig:PredMAP}, we find the uncertainty to be systematically lower when predictions are based on longer observational periods. The longer the available observation period, the narrower the posterior and the smaller the difference between the two types of predictions and their respective uncertainty in the $p_k$-space.

    \section{Discussion}
         \label{sec:discussion}

We have explored a modelling and inference framework for forecasting SIS epidemics spreading on networks. The surrogate model is based on a BD process. The effect of the contact structure has been condensed into a birth-rate parameter, which is proportional to the average number of SI-links for a given number of infected nodes. Our empirical validation has confirmed that the BD model is well suited to describe the evolution of an SIS epidemic on a network (Figs.~\ref{fig:bias} d-f, \ref{fig:iqd} d-f). Both the expectation and the intrinsic stochasticity of the epidemic trajectories are well reproduced even though our model formulation contains a mean-field approximation. The parametric model for the number of SI-links, which has been introduced to enable the inference of network class and model parameters, is suitable for the range $50 \lessapprox k \lessapprox 950$ (Fig.~\ref{fig:ak_curves}). Hence, simulations with the BD model with the parametric $a_k(C,\alpha,p)$ should not be initialised with fewer than $50$ infected nodes (Fig.~\ref{fig:bias} a-c,~\ref{fig:iqd} a-c).

Network class and epidemic parameters can be reliably inferred when observations are available from an early stage of the epidemic up to the quasi-steady state. However, in realistic prediction scenarios, observations are only available up to the current state of the epidemic. The accuracy of the network class inference is sensitive to the observational time span (Fig.~\ref{fig:ClassSens}). Uncertainty increases as observational time span is reduced. This is because epidemics, though spreading on networks from distinct classes, can exhibit very similar trajectories through their earlier stages and only diverge when approaching the quasi-steady state (Fig.~\ref{fig:trajcetories}). As discussed in \cite{allen2021predicting} for instance, the uncertainty of the future course of an emerging epidemic during its early stages is dominated by the intrinsic stochasticity of disease transmission. It is thus no surprise that observations from an early stage of the epidemic appear not to contain sufficient information about the network class. 

In predictions based on observations up to and initialised at $I=160$, the prediction uncertainty is dominated by the uncertainty of the parameters/network classes. In predictions based on observations up to and initialised at $I=160$, the prediction uncertainty stems in about equal proportions from parameter/network class uncertainty and intrinsic stochasticity of the epidemic spreading (Fig.~\ref{fig:PredUncer}). Thus, and especially for shorter observational periods and hence predictions initialised early during the epidemic, considering parameter uncertainty is crucial for providing meaningful information about prediction uncertainty \citep[see also][]{castro2020turning, wilke2020predicting}. The results suggest that for most cases the credible intervals obtained provide reliable uncertainty information for the epidemic forecasts (see also the Supplementary Material). If longer observational time spans are available, point estimate-based predictions are potentially useful as well (Fig.~\ref{fig:PredMAP}). 

Our study differs from other approaches in network inference in so far as our aim here is not to infer the existence, or otherwise, of links but rather to infer the most likely network class that led to the observed population-level data resulting from an epidemic spreading on it. As a result, the data needed for inference does not contain node- or link-level information. There are both advantages and disadvantages to such an approach. On the one hand, the computation of the likelihood in our case is more straightforward and the data needed for inference is modest. On the other hand, if more detailed data is available, the proposed model will not be able to capture it nor benefit from it. However, more complex models will need large quantities of detailed data (i.e., in the case of cascades, the data needs to contain cascades starting from, or involving, as many nodes as possible, \citet{gomez2012inferring}) to produce acceptable results with large computational burden. The choice of model and inference will depend on the context.

There are many directions in which the current model and inference scheme can be developed. First, we only explored three network classes where the key difference was degree heterogeneity. However, networks displaying degree-degree correlations, clustering, spatial structure or some type of meso-scale structure, such as communities, may be of interest as they are more representative of real-world scenarios. Equally, from a theoretical viewpoint, lattices could be considered. This is a non-trivial task and depending on which network property or combination of properties we choose to model, it may turn out that the birth-rates of the BD process will no longer be parabola-like and the proposed parametric $a_k$-model may no longer provide a satisfactory fit. However, we expect that more complex models will be able to capture the birth-rates in the BD process resulting from such more exotic networks. Another natural extension would be to consider more complex epidemic models, such as SIR, where the corresponding BD model will now have $O(N^2)$ equations and the birth-rates of the BD process will define a surface rather than a curve. However, and perhaps more interestingly, the excellent agreement between the exact and surrogate model, leads us to believe that a rigorous proof that quantifies the error between the exact and BD models may be possible. For example, it is clear that as a Regular network becomes more densely connected, and in the limit of number of links going to $N-1$, the BD model becomes exact.

\section*{Acknowledgements}
All authors acknowledge support from the Leverhulme Trust through the Research Project Grant RPG-2017-370.

\bibliographystyle{apalike} 
\bibliography{bib}

\end{document}